\documentclass[sigconf]{acmart}
\usepackage[utf8]{inputenc}

\usepackage{xcolor}
\usepackage{xspace}
\usepackage{url}
\usepackage{textcomp}
\usepackage{breakurl} 
\usepackage{listings}
\usepackage{cleveref}
\usepackage{algorithm}

\usepackage{balance}

\newcommand{\gptBase}{\texttt{gpt-3.5}\xspace}
\newcommand{\gptFour}{\texttt{gpt-4}\xspace}

\begin{document}

\title{One Step at a Time: Combining LLMs and Static Analysis to Generate Next-Step Hints for Programming Tasks}

\author{Anastasiia Birillo}
\affiliation{
  \institution{\textit{JetBrains Research}}
  \city{Belgrade}
  \country{Serbia}
}
\email{anastasia.birillo@jetbrains.com}

\author{Elizaveta Artser}
\affiliation{
  \institution{\textit{JetBrains Research}}
  \city{Munich}
  \country{Germany}
}
\email{elizaveta.artser@jetbrains.com}

\author{Anna Potriasaeva}
\affiliation{
  \institution{\textit{JetBrains Research}}
  \city{Belgrade}
  \country{Serbia}
}
\email{anna.potriasaeva@jetbrains.com}

\author{Ilya Vlasov}
\affiliation{
  \institution{\textit{JetBrains Research}}
  \city{Belgrade}
  \country{Serbia}
}
\email{ilya.vlasov@jetbrains.com}

\author{Katsiaryna Dzialets}
\affiliation{
  \institution{\textit{JetBrains}}
  \city{Munich}
  \country{Germany}
}
\email{katsiaryna.dzialets@jetbrains.com}

\author{Yaroslav Golubev}
\affiliation{
  \institution{\textit{JetBrains Research}}
  \city{Belgrade}
  \country{Serbia}
}
\email{yaroslav.golubev@jetbrains.com}

\author{Igor Gerasimov}
\affiliation{
  \institution{\textit{JetBrains}}
  \city{Berlin}
  \country{Germany}
}
\email{igor.gerasimov@jetbrains.com}

\author{Hieke Keuning}
\affiliation{
  \institution{\textit{Utrecht University}}
  \city{Utrecht}
  \country{Netherlands}
}
\email{h.w.keuning@uu.nl}

\author{Timofey Bryksin}
\affiliation{
  \institution{\textit{JetBrains Research}}
  \city{Limassol}
  \country{Cyprus}
}
\email{timofey.bryksin@jetbrains.com}

\renewcommand{\shortauthors}{Birillo, Artser, Potriasaeva, Vlasov, Dzialets, et al.}

\begin{abstract}
Students often struggle with solving programming problems when learning to code, especially when they have to do it online, with one of the most common disadvantages of working online being the lack of personalized help. This help can be provided as \textit{next-step hint generation}, \textit{i.e.}, showing a student what specific small step they need to do next to get to the correct solution. There are many ways to generate such hints, with large language models (LLMs) being among the most actively studied right now.

While LLMs constitute a promising technology for providing personalized help, combining them with other techniques, such as static analysis, can significantly improve the output quality.
In this work, we utilize this idea and propose a novel system to provide both textual and code hints for programming tasks. The pipeline of the proposed approach uses a chain-of-thought prompting technique and consists of three distinct steps: (1) generating \textit{subgoals} --- a list of actions to proceed with the task from the current student's solution, (2) generating the \textit{code} to achieve the next subgoal, and (3) generating the \textit{text} to describe this needed action. 
During the second step, we apply static analysis to the generated code to control its size and quality. 
The tool is implemented as a modification to the open-source JetBrains Academy plugin, supporting students in their in-IDE courses.

To evaluate our approach, we propose a list of criteria for all steps in our pipeline and conduct two rounds of expert validation.
Finally, we evaluate the next-step hints in a classroom with 14 students from two universities. Our results show that both forms of the hints --- textual and code --- were helpful for the students, and the proposed system helped them to proceed with the coding tasks.

\end{abstract}



\maketitle

\section{Introduction}\label{sec:introduction}

A popular format for learning programming is Massive Open Online Courses (MOOCs)~\cite{irwanto2023massive}, which became particularly popular during the pandemic~\cite{cramarenco2023student}. 
However, while many studies show that personalized feedback helps students learn the material and solve problems faster~\cite{liffiton2023codehelp, rivers2013automatic, xiao2024exploring, deeva2021review}, it is hard to provide personalized help in MOOCs because of the large number of students. 
To provide this help, the process needs to be automated, with many approaches to this being researched~\cite{keuning2018systematic}.
One way to provide help when working on a problem is \textit{next-step hint generation}~\cite{keuning2018systematic},
the purpose of which is to suggest the next \textit{step} for the student's solution, leading them to the final solution that passes all tests.

In the era of Large Language Models (LLMs), many works have explored the possibility of applying LLMs in the context of feedback generation~\cite{roest2024next, liu2024teaching, liffiton2023codehelp, aruleba2023integrating, kiesler2023exploring}. 
The authors of such works mainly use the open API of popular LLMs such as \gptBase~\cite{kazemitabaar2024codeaid, roest2024next, liffiton2023codehelp} and \gptFour~\cite{liu2024teaching}, and utilize various \textit{prompt engineering} techniques~\cite{marvin2023prompt}. 
Some works focus on tasks for beginners in Python~\cite{roest2024next} or C~\cite{kazemitabaar2024codeaid}, but the majority offer universal approaches that are language-independent~\cite{liu2024teaching, liffiton2023codehelp}. 
A next-step hint can be generated in different forms, from a simple text message~\cite{roest2024next} or the solution in (pseudo) code~\cite{kazemitabaar2024codeaid} to a combination of both forms~\cite{xiao2024exploring}.
What all these studies have in common is that they use LLMs directly, without any additional processing to check the LLM output. 
This could affect the accuracy of the results, as LLMs often give incorrect output due to hallucinations, limited context, and probabilistic nature~\cite{liu2024exploring, mcintosh2023culturally}.

Recently, in-IDE learning has been described as a new possible learning format for MOOCs~\cite{birillo2024bridging}. 
The main purpose of this approach is to integrate the learning process with professional Integrated Development Environments (IDEs), helping students learn not only how to program but also how to use IDE features when coding.
The format was implemented as an open-source JetBrains Academy plugin~\cite{jetbrains-academy-plugin} --- an extension of JetBrains IDEs, such as IntelliJ IDEA~\cite{intellijIdea}, PyCharm~\cite{pycharm}, or CLion~\cite{cilion}.
However, this learning format does not currently offer personalized help to students.

In this work, we present a new approach to next-step hint generation that combines the power of static analysis and state-of-the-art LLMs to achieve better hint quality. 
We applied the chain-of-thought prompting approach~\cite{wei2022chain}, which splits the prompt into multiple smaller prompts chained together. The final pipeline consists of three steps: (1)~generating \textit{subgoals} --- a list of actions to proceed with the task from the current student's solution, (2) generating the \textit{code} to achieve the next subgoal, and (3) generating the \textit{text} to describe this needed action. 
Since recent research has shown that providing just one level of help is not enough~\cite{xiao2024exploring}, we provide help in two forms --- textual hints and code hints --- with the student choosing the desired type. 
During code generation, we employ static analysis to control the size of the hint using several heuristics, as well as its code quality --- using inspections~\cite{keuning2017code}.
In the proposed approach, we use in-IDE static analysis, but it can be replaced with any other static analysis tools outside of the IDE.

To evaluate the proposed hint system, we implemented it as an extension of the JetBrains Academy plugin, since it is open-source and publicly available.
We conducted two rounds of expert validation to ensure the system's quality based on the proposed lists of criteria for the different steps in our pipeline.
Finally, we evaluated the next-step hints in a classroom with 14 students from two universities, demonstrating their usefulness. 

The rest of the paper is organized as follows. Section~\ref{sec:background} describes the related work and the in-IDE learning format. Section~\ref{sec:approach} describes how the proposed hint system works from the perspective of the students, while Section~\ref{sec:approach:internal} describes how it works under the hood. Section~\ref{sec:design} explains how we designed the system and presents the internal validation that we used to achieve high quality, and Section~\ref{sec:evaluation} describes the evaluation on students. Finally, Section~\ref{sec:threats} identifies possible limitations of this study, and Section~\ref{sec:conclusion} concludes the work.

\section{Background}\label{sec:background}

 \begin{figure*}[t]
    \centering
    \includegraphics[width=\linewidth]{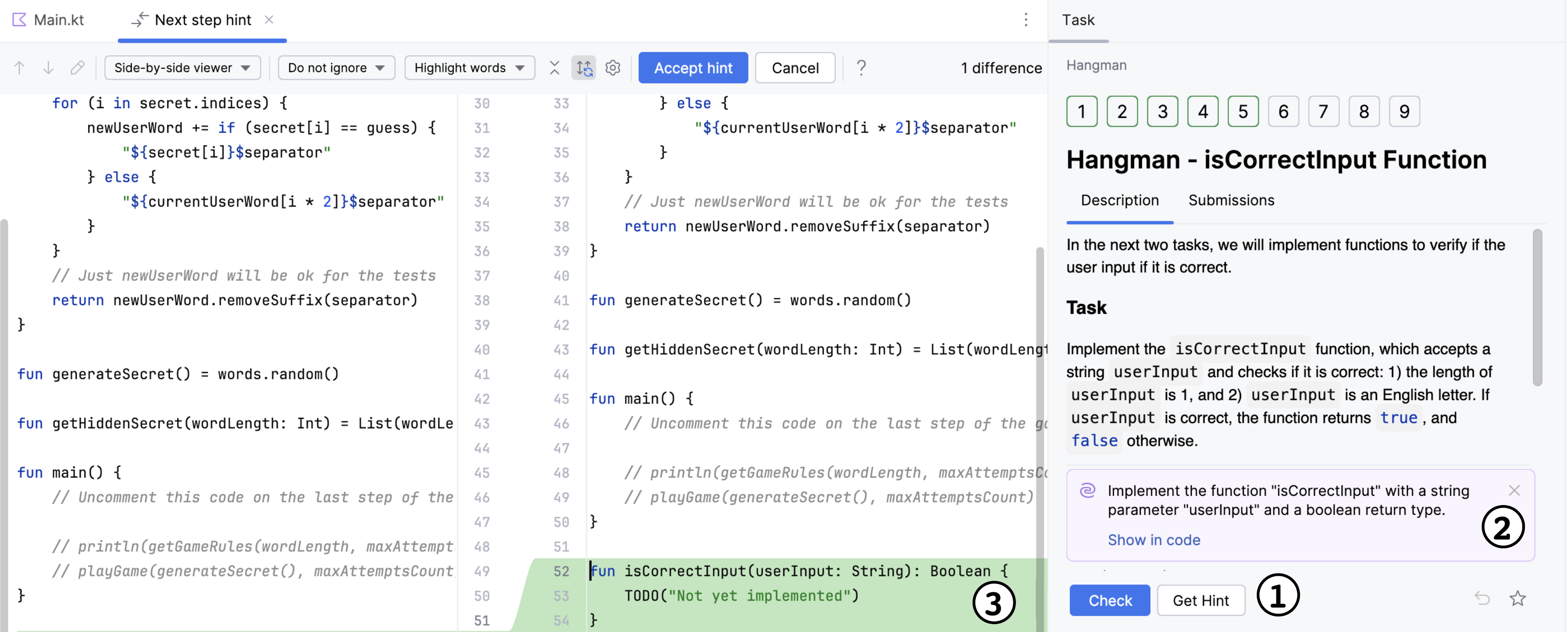}
    \caption{The UI of the JetBrains Academy plugin together with an example of the developed next-step hint: (1) the ``Get hint'' button, (2) textual hint, and (3) code hint.}
    \label{fig:final_hint}
    \vspace{0.5cm}
\end{figure*}

\subsection{Personalized Feedback}

Feedback generation is a common way for providing personalized help to students during the learning process~\cite{liffiton2023codehelp, rivers2013automatic, xiao2024exploring, deeva2021review}. One of the possible ways to provide such help is \textit{next-step hint generation}, \textit{i.e.}, showing a student what specific small step they need to do next. There are many ways to generate such hints~\cite{keuning2018systematic}, from predefined rules or templates~\cite{its2017haskell} to using systems based on previous student submissions~\cite{rivers2013automatic, its2017isnap} or large language models (LLMs)~\cite{roest2024next, liu2024teaching, liffiton2023codehelp, aruleba2023integrating} that have recently become popular.
LLM-based systems allow providing immediate feedback to a lot of students without involving the teacher and, compared to data-driven approaches, do not require gathering a large amount of data on the student submissions. This section provides an overview of recent approaches to generating next-step hints with LLMs.

\textit{StAP-tutor}~\cite{roest2024next} is a web-based application that gives students a textual next-step hint in case they get stuck when solving programming problems.
The tool uses the \gptBase model under the hood and provides hints for Python programs at the click of a hint button.
The approach focuses on a small number of simple exercises and has not been tested with more complex tasks.
The other disadvantage of this tool is that it only provides a textual hint, whereas according to a recent study~\cite{xiao2024exploring}, students might need several levels of hints for better understanding.

Recently, Liffiton et al. developed \textit{CodeHelp}~\cite{liffiton2023codehelp}.
Like \textit{StAP-tutor}, this tool only generates textual hints, but in this case, students need to write their own questions to prompt the LLM and provide the context information in a special form to get help.
The tool separates the student input into a code part, an error message, and the student question to build the final prompt for the LLM in a more structured way under the hood.
The response consists of a detailed text explanation of the next step that the student should proceed with.
The authors use a combination of the \gptBase and \texttt{davinci} models for output processing to remove any code from the resulting hints.
They also added a sufficiency check to notify students in case their prompt is too vague.
One of the main drawbacks is that students in this system have to formulate their own specific questions, which can be challenging, especially for novices~\cite{knoth2024ai}.

\textit{CodeAid}~\cite{kazemitabaar2024codeaid} is a web-based application that allows students to not only find mistakes in the code, but also answer general questions and explain the code.
Similar to the previous tools, the hint system utilizes the \gptBase model and provides hints for programs in C.
The hints themselves are generated as text and pseudocode to support students in transitioning from understanding concepts to independently writing their code.
In the paper, the authors noted that first generating the correct code and only then the textual hint greatly improves quality and accuracy. 
The main disadvantages of this tool are that it can be used only with simple tasks and that students must write their own questions.

Unlike the above-mentioned tools, \textit{CS50.ai}~\cite{liu2024teaching}, introduced by Liu et al., is not only a web-based application but also has a plugin integrated with Visual Studio Code~\cite{vscode}. 
Getting help inside the IDE allows students to learn in a production environment that they will later face in their work.
To improve the quality of hints, the authors use a more powerful \gptFour model along with retrieval-augmented generation (RAG)~\cite{lewis2020retrieval}, which reduces LLM hallucinations by introducing new information to the LLM from external sources.
As a hint, the authors provide the students with a list of step-by-step actions that should be taken to solve the task.
Like the previous tool, \textit{CS50.ai} offers some additional features such as explaining code and checking code style.
The authors implemented a security mechanism to prevent the system's abuse, but the tool still requires students to write their own questions as prompts.

In summary, the described works have tried different AI-based techniques to provide students with the next-step hints in different forms --- textual or textual together with code.
As a recent study has shown, the combination of different forms can be crucial for the students' understanding~\cite{xiao2024exploring} and should thus be used.
One of the biggest challenges in these works is related to the use of LLMs, as they often produce hallucinations and may ignore prompt instructions~\cite{liu2024exploring, mcintosh2023culturally}, which can directly affect the quality of the hints. This work addresses this challenge and uses static analysis as a post-processing step to control hint size and code quality.
Finally, the chain-of-thought prompting technique~\cite{wei2022chain} could help increase the accuracy of the generated hints, but this technique has not yet been used in this type of systems.

\subsection{In-IDE Learning}

\begin{figure*}[t]
    \centering
    \includegraphics[width=\linewidth]{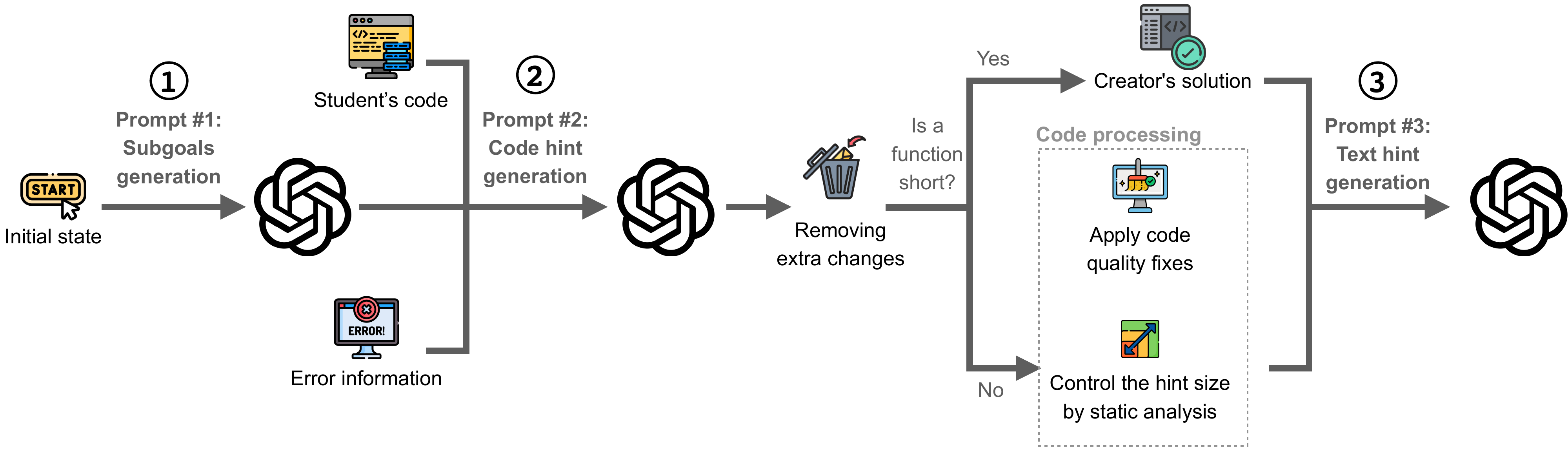}

    \caption{The overall pipeline of the next-step hint system.}

    \label{fig:generalPipeline} 
\end{figure*}

Recently, the new \textit{in-IDE learning format} was presented~\cite{birillo2024bridging} in the form of the open-source JetBrains Academy plugin~\cite{jetbrains-academy-plugin}, an extension for JetBrains IDEs that allows studying programming using a professional environment.
The plugin aims to bring a more realistic experience to those practicing their programming skills~\cite{birillo2024bridging}.
There are already over 40 available courses in this format, with all of them being free and any third-party educator being able to create their own course with this format.
A course consists of theoretical and practical \textit{tasks}. A theoretical task explains a concept in a text or a video form, while a practical task can be either a quiz or a programming assignment to be evaluated by a testing system.
The tasks can be isolated from each other or form a single larger \textit{project} where students have to write the code step-by-step. You can see an example of such project in Figure~\ref{fig:final_hint}, with the numbers in the top-right corner representing nine step-by-step tasks that the students solve to complete the project.
All of this makes the in-IDE learning format a good environment for implementing a novel next-step hint system: the programming assignments can be complex, and the IDE conveniently provides several crucial features, such as automated tests and an API for static analysis.

\section{The Next-Step Hint System: Student's Perspective}\label{sec:approach}

In this section, we describe how the proposed \textit{next-step hint system} works from the perspective of a student. In this work, we focused on console applications as target tasks, which require not only coding the main logic of the program by implementing several related functions, but also implementing some helper functions like reading and handling the user's input. Such tasks are inherently more complex, they represent a more realistic studying environment but may also require more help for the student. We target the tasks in the Kotlin language, because Kotlin courses are among the most popular in the JetBrains Academy plugin~\cite{jetbrains-academy-marketplace}.

You can see all the main elements of the next-step hint system in Figure~\ref{fig:final_hint}. The entry point for the interaction with the system is a ``Get Hint'' button, which was added to the task panel of the existing interface (see \textbf{(1)} in Figure~\ref{fig:final_hint}). 
Pressing the button initiates the generation of the next-step hint, which is provided in two forms --- \textit{textual} and \textit{code}.
The textual hint covers two levels of help considered by Xiao et al.~\cite{xiao2024exploring}: (a) \textit{instrumental} help, which informs students on what to do next in concise, descriptive sentences by showing it in a purple window near the ``Get Hint'' button (see \textbf{(2)} in Figure~\ref{fig:final_hint}); and (b) \textit{orientational} help, which informs students where they should focus their attention by highlighting the position in the code editor.
The textual hint also includes a ``Show in code'' button (see \textbf{(2)} in Figure~\ref{fig:final_hint}), which provides access to the code hint.
Pressing this button opens a code-diff window --- a window that displays the changes between two versions of source code (the current student code and the code after applying changes from the code hint), highlighting additions, deletions, and modifications for easier comparison (see \textbf{(3)} in Figure~\ref{fig:final_hint}). 
This help represents the \textit{bottom-out} help~\cite{xiao2024exploring}, which shows students the exact code they need to write for the next step.
The code-diff window also has two action buttons: an ``Accept'' button to accept the code hint and automatically apply the suggested changes to the student's code and a ``Cancel'' button to cancel the code hint.

Students may request a next-step hint at any time during the task, except for when it is already solved. 
Since JetBrains Academy is a MOOC platform where students engage and take courses voluntarily~\cite{birillo2024bridging}, we did not consider limiting the number of hint requests in the current version of the next-step hint system.

\section{The Next-Step Hint System: Internal Design}\label{sec:approach:internal}

In this section, we describe how the proposed system works under the hood. The system uses \gptFour for all LLM interactions described further in this section.
The details of \textit{why} we designed the system this way and \textit{what students think} about the hints are discussed in Sections~\ref{sec:design} ~and~\ref{sec:evaluation}, respectively. 

\subsection{General Pipeline}

The general pipeline of the proposed next-step hint system is presented in Figure~\ref{fig:generalPipeline} and consists of three main stages: \textbf{(1)}~generating \textit{subgoals}, \textbf{(2)} generating the \textit{code hint}, and \textbf{(3)} generating the \textit{textual hint}. 
The stage of generating subgoals aims to analyze the task and the current student's code, determine the path to the final solution, and break it down into a list of coding-related actions with sufficient granularity --- subgoals. 
The purpose of generating code and textual hints is to create the corresponding hints, taking into account the task's subgoals and the current state of the student's solution. Let us consider each stage in more detail.

\subsection{Generating Subgoals}\label{sec:approach:subgoals}

Large language models are probabilistic in nature and may ignore some parts of prompts in their output~\cite{webson2021prompt}, which may lead to generating hints that contain multiple next steps at once even when asked not to.
To mitigate this effect, we introduced the generation of \textit{subgoals}, which strives to break down the task into smaller, manageable steps~\cite{margulieux2020effect}. 
The key idea behind generating subgoals is to analyze the task in general first before generating a next-step hint, which depends on the specific state of the student's solution. 
This approach ensures that the provided hints are finely granulated, thus maintaining an optimal level of challenge and support.
A similar approach, showing a list of next-step actions as a hint, was recently presented by Liu et al.~\cite{liu2024teaching} and showed good performance. 

The final prompt for generating subgoals can be found in the supplementary materials~\cite{suplementary}. 
We prompt the LLM to generate an ordered list of steps to solve the given task, providing six parameters with specific details about the task:
\begin{itemize}
    \item \textbf{Task description}.
    \item \textbf{Set of signatures of functions that might be implemented}, which is extracted from a model solution (\textit{i.e.}, solution provided by the task creator) that is not visible to the student. We do not provide the entire model solution to the LLM to avoid bias towards the approach suggested by the task creator~\cite{keuning2018systematic}.
    \item \textbf{Set of existing functions within the student's code} that have already been implemented in the previous tasks of the larger project and can thus be used in the solution. These functions are mentioned separately from the functions of the model solution so that the LLM can directly use the already implemented ones.
    \item \textbf{Static predefined hints}, optionally provided by the task's creator. These are intended to help students with common difficulties, but they are not dynamic and do not depend on the student's solution. 
    \item \textbf{List of theory topics introduced in the current project}, such as "variables", "loops", or "functions", to ensure that the generated subgoals are aligned with the covered topics. This information is extracted from the theoretical tasks in the project, the names of which correspond to the coding concepts that they teach.
    \item \textbf{Set of string literals extracted from the model solution}, which the students might use in their solutions. We added this because we observed that GPT-like models often generated strings close to the strings from the task description but slightly rephrased. Since we focused on console applications, using the exact string constants is often crucial for completing the task.
\end{itemize}

To force the LLM to use built-in functions from the standard libraries of the required programming language, we explicitly specify the language that should be used to solve the coding problem.

During our experiments with prompting, we also observed that the LLM often tended to make broad subgoals, not achieving the desired granularity.
We try to make each subgoal responsible for an independent small coding step, equivalent to a next-step hint.
To mitigate this broadness, we also specify the minimum number of subgoals to generate --- at least 6. 
Introducing a lower bound for the number of subgoals led to some of the generated subgoals being non-code-related, especially on the most straightforward tasks that require only a few steps to complete. 
For example, the LLM suggested to "read the task carefully", "save the file," or "run the solution" as subgoals. 
To avoid such subgoals being proposed as a next-step hint, we ask the LLM to mark each subgoal as a "code" or "no-code" step, and then additionally remove any "no-code" steps from the final list as a post-processing of the LLM response.
As a result of this stage, we obtain a list of granular subgoals.

\subsection{Generating Code and Textual Hints}\label{sec:approach:hints}

The next stage in our approach is generating the code hint and the textual hint (see \textbf{(2)} and \textbf{(3)} in Figure~\ref{fig:generalPipeline}). 
The code hint is generated first, followed by the textual hint (even though they are shown to the students in the reverse order).
According to Kazemitabaar et al. ~\cite{kazemitabaar2024codeaid}, generating the correct code for the next step initially, followed by the pseudo-code, proved to deliver significantly better results.
Upon comparing the validation results for different orders of generating hints, we reached the same conclusion (see Section~\ref{sec:design} for more details).

\subsubsection{Code hint prompt}

The final prompt for the code hint can be found in the supplementary materials~\cite{suplementary}.
We prompt to generate a modified version of the student's code by submitting the generated \textbf{list of subgoals} of the corresponding task and the \textbf{student's code}. If the student already tried to run the tests for the current solution, we also provided the LLM with the \textbf{reported errors}, which was demonstrated to enhance the quality of the LLM output~\cite{liffiton2023codehelp}. 

\subsubsection{Code hint quality improvement}

One of the main features of our approach is that we use \textit{static analysis} to control the generated code provided by the LLM and improve the resulting hint.

\textbf{Removing extra changes}. To generate a focused and relevant hint, we omit irrelevant suggestions, such as when the LLM tries to improve the quality of the code in functions that the student already implemented in the previous tasks of the larger project.
To achieve this, we compare the current student solution with the model solution and extract functions that should be added or changed. We ignore suggested changes in other functions, and if the LLM proposes to change several relevant functions, we only keep the changes for the first one.

\textbf{Handling short functions}. 
Now that we selected the specific function to which the hint should be applied, we check whether this function is short or long. 
We believe that very short functions already correspond to an appropriate size for one next-step hint and do not require further control.
Moreover, for short functions we propose using the model solution provided by the task creator instead of the LLM output, as it is already suitable. 
We treat short functions differently because of the inherent unreliability of LLMs due to their non-deterministic output and the potential for hallucination~\cite{liu2024exploring, mcintosh2023culturally}.
The threshold for defining a function as short has been empirically determined to be not more than three lines in its body.
If the changed function is longer than this, it might represent a complex step and thus follow a different approach from the one in the model solution, which is why in this case we use LLM-generated code that takes into account all the aspects of the current student solution. However, for the hint to be concise and to provide code with good quality, the LLM-generated code requires further control by static analysis, which we describe below.

\textbf{Ensuring code quality}. 
One of the crucial aspects of the code being shown to students is its quality~\cite{keuning2017code}, however, the code generated by LLMs may lack in this regard~\cite{liu2024no, tambon2024bugs}.
To improve the quality of the generated code, we utilize the IDE's code quality inspections that can be used for optimizing code and correcting common security and style issues.
We selected 30 Kotlin inspections, for which automatic fixes are available.
For instance, one inspection transforms a comparison into a Kotlin range if possible, \textit{e.g.}, \textit{month >= 1 \&\& month <= 12} into \textit{1..12}.

\begin{figure*}[t]
    \centering
    \includegraphics[width=\linewidth]{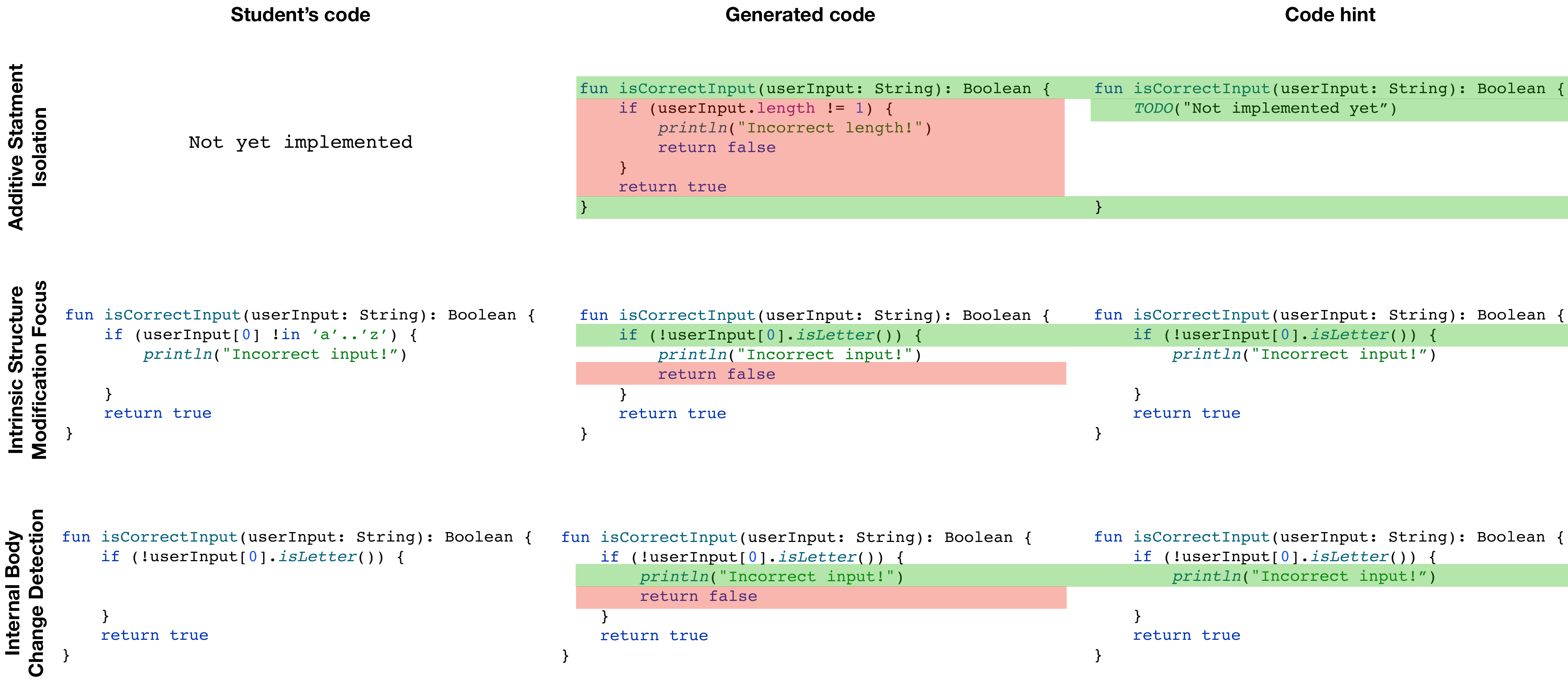}
    \vspace{-0.5cm}
    \caption{Illustration of the impact of applying heuristics to reduce changes in student code. The green highlighting indicates the changes that were kept for the code hint. The red highlighting indicates changes that were removed by the heuristic.}

    \label{fig:heuristics}
\end{figure*}

\textbf{Controlling hint size}. Finally, LLMs do not always comprehend that only one step needs to be generated, even within one function. We believe that a good hint provides one logical action, \textit{e.g.}, create a function, use a \textit{for} loop, etc., whereas the generated code often contains several such steps, \textit{e.g.}, generating a function or a \textit{for} loop with the full body.
Consequently, we decided to control this process ourselves with static analysis from the IDE, however, it can be replaced with any other static analysis tool.
We developed three general heuristics, each applied to six control structures. You can find the full list of heuristics with detailed descriptions in the supplementary materials~\cite{suplementary}, and here we will consider specific motivating examples presented in Figure~\ref{fig:heuristics}. 
The first row illustrates the case when the LLM returned a new function to be implemented. 
Subsequently, we apply the heuristic \textit{Additive Statement Isolation}, resulting in the function definition being retained, while the body is replaced with a \textit{TODO} expression. 
The same approach can be applied to other newly added constructs. 
The second row illustrates that several lines within the \textit{if} construct were simultaneously changed in the generated code, involving both the condition and the body. 
After applying the \textit{Intrinsic Structure Modification Focus} heuristic, we only keep the difference related to the condition and remove the remaining ones. 
In the last row, it can be again observed that several lines of code have been changed in the \textit{if} construct, but this time all of them in the body. 
Subsequently, the heuristic \textit{Internal Body Change Detection} was applied, resulting in only keeping the first modification (the \textit{print} expression). 

Applying these heuristics, coupled with the initial generation of subgoals, allows us to make sure that the proposed hint really does represent a single granular step towards the correct solution.

\subsubsection{Textual hint prompt}

The final step of our pipeline is generating the textual hint (see \textbf{(3)} in Figure~\ref{fig:generalPipeline}). Even though the textual hint is shown to the students first, it is generated after the code hint, as this generation order improves the overall quality of the hints~\cite{kazemitabaar2024codeaid}. The final prompt for the textual hint can be found in the supplementary materials~\cite{suplementary}. 
We prompt to generate a textual hint based on the given \textbf{current  student's code} and an improved version of the \textbf{code generated at the previous step}. 
We instruct the LLM to provide a brief textual instruction in an imperative form in the response, without including explanations and code into the resulting textual hint. 
\section{Internal Validation}\label{sec:design}

\subsection{Overview}

This section provides an overview of the research process we carried out to develop our solution. Section~\ref{sec:design:ux} describes a pilot UX study with students, the main goal of which was to find a convenient and efficient way to show the next-step hints.  Section~\ref{sec:design:subgoals} describes the process of designing and validating the prompt for generating subgoals. Section~\ref{sec:design:hints} describes the process of designing and validating the prompts for generating textual and code hints.

All validation data in this section is based on the ``Kotlin Onboarding: Introduction'' course~\cite{kotlin-onboarding-introduction}, a Kotlin course for beginners in the in-IDE learning format. The course covers basic programming concepts, such as variables, conditional operators, loops, and functions. The course consists of six console projects, the description of which can be found in the supplementary materials~\cite{suplementary}. In total, the six projects contain 50 individual coding tasks.

For financial reasons, we used \gptBase for all experiments in this section, even though \gptFour was used in the final version. Running all the intermediate queries for dozens of tasks and multiple rounds of validation resulted in a lot of requests. More importantly, these experiments were aimed at finding high-level problems with prompts and fixing them. A recent study~\cite{prather2023robots} demonstrated how much better \gptFour performs in solving programming assignments, which is why we used it in the final version of our system. Section~\ref{sec:evaluation} describes the evaluation on students that used the final version with \gptFour, and we leave more detailed comparison of different models for future work.

\subsection{UX Design of The Next-Step Hint System}\label{sec:design:ux}

\begin{figure*}
    \centering
    \includegraphics[width=\linewidth]{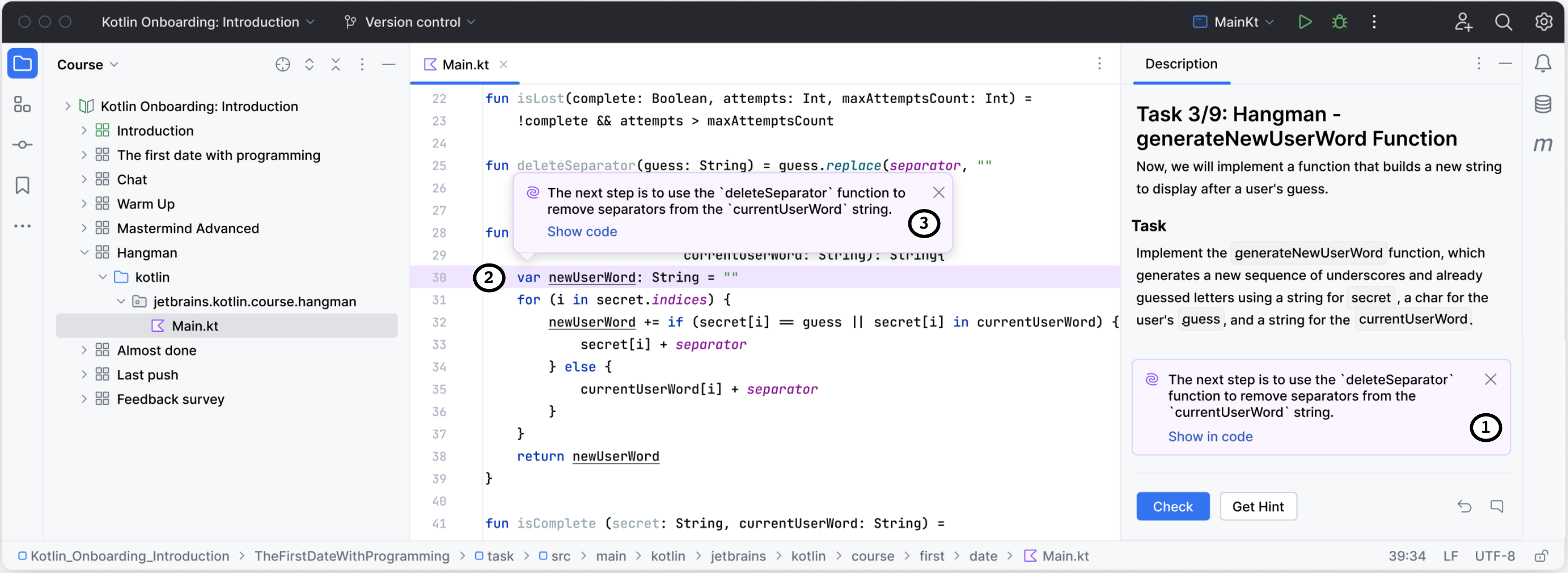}
    \caption{Proposed UX options. \textbf{Prototype A}: a textual hint is shown in the \textit{Task description} panel (1). \textbf{Prototype B}: a textual hint is shown in the \textit{Task description} panel, and the line is highlighted where changes have to be made (1 + 2). \textbf{Prototype C}: a textual hint is shown in the code editor and points to the location where changes have to be made (3).}
    \label{fig:ux_options}
\end{figure*}

In the proposed next-step hint system, students are first shown a textual hint, after which they may press a button to view the code-diff window and apply the hint automatically. Since the IDE setting is inherently complex and the in-IDE learning format is not yet well studied~\cite{birillo2024bridging}, it was not obvious to us how we should show hints to students. Therefore, we conducted a user experience (UX) study to identify a better solution that would work for different tasks and for students with different levels of experience.  We did not focus on the code hint, since the IDE already provides a good mechanism for code comparison that could be reused for our system, and thus focused our attention on how to display the textual hint. To compare different options, we conducted 15-minute comparative usability interviews with nine students. We selected this method because it was simple yet highly informative~\cite{van2007testing}. The rest of the section describes in detail the options proposed to students, the conducted study, and its results.

\subsubsection{Approaches to displaying a textual hint} We considered three options to display a textual hint (see Figure~\ref{fig:ux_options}):
\begin{itemize}
    \item \textbf{Prototype A: Hint in the same context with the task}. Element \textbf{(1)} shows the hint in the same context window as the task description. The assumption for this prototype was that it is convenient to keep all the information in the same place: the task, the button, and the response.
    \item \textbf{Prototype B: Hint in the same context with the task + highlighting the position}. Element \textbf{(1)} together with \textbf{(2)} shows the second option, not only keeping the student in context with the task but also helping them find the place where the hint should be applied. The main assumption is that students may have problems with navigating the code, especially if there are several functions in the solution.
    \item \textbf{Prototype C: Hint directly at the position}. Element \textbf{(3)} shows the third option,  which separates the task and the hint system and shows them independently. The idea is to draw the student's attention to the code, thus helping them solve the problem.
\end{itemize}

\subsubsection{Interview design} We developed interactive Figma~\cite{figma} prototypes for all three approaches on three different tasks: a simple \textit{"Hello, world"} example, adding a new function to the solution, and resolving an error in the solution with several functions. The participants interacted with the prototypes and provided feedback in real-time. Having tried all three options, each participant indicated the one they found the most convenient and the reasons for this.

\subsubsection{Interview participants} A total of nine students from two universities participated in the study. The age range of the students was 18 to 23 years old. Seven students are getting a Bachelor's degree, while two students are getting a Master's degree.
Six students self-reported as experienced in programming, while three self-reported as novices. The programming languages that the students were familiar with included Kotlin, C++, Java, and Python. Six students had previous experience with the JetBrains Academy Plugin and the in-IDE learning format, while three students had little to no such experience.

\subsubsection{Interview results} The results of the usability session with nine students indicated that \textbf{Prototype B} (Figure~\ref{fig:ux_options}, \textbf{(1)} together with \textbf{(2)}) was the most preferred, with six students selecting it.  \textbf{Prototype C} (Figure~\ref{fig:ux_options}, \textbf{(3)}) received two votes, and \textbf{Prototype A} (Figure~\ref{fig:ux_options}, \textbf{(1)}) was the least preferred, with only one student selecting it.
Among the key insights, we found that highlighting the position in the code where changes need to be applied is crucial. Regarding \textbf{Prototype C}, the main problem for the students was that the code overlapped with the hint panel, and the students could not read some of the functions. Also, this design does not allow keeping the textual hint together with the code hint, which was confusing for the students. Based on these results, we selected \textbf{Prototype B} for the final version. 

Having selected the preferred UX, we moved on to the expert validation of different stages of our pipeline, which involved designing the validation criteria based on the existing research and our experience, as well as conducting two rounds of manual expert labeling. The following two sections describe this process in detail.

\subsection{Validating the Generation of Subgoals}\label{sec:design:subgoals}

As the first part of the validation, we carried out the expert validation for the generation of subgoals.

\subsubsection{Validation criteria}

Based on prior work~\cite{decker2019using, margulieux2019design} and our experience, we propose 8 criteria to validate the subgoal generation. 

\begin{itemize}
    \item \textbf{Amount} criterion evaluates the total number of subgoals generated for a given task.
    \item \textbf{Specifics} criterion assesses whether the subgoals are related to specific actions, such as creating variables, calling functions, or using coding constructs like loops or \textit{if} statements, and not something broad like repeating the task statement.
    \item \textbf{Independence} criterion checks that subgoals do not refer to each other and are independent, since we do not show the list of subgoals in the final hint and the student will not understand the reference.
    \item \textbf{Coding-specific} criterion checks if all the subgoals are related directly to coding tasks, with a poor example being "Think about the problem". Even though we asked the LLM to mark non-code-specific actions and then removed them, the LLM sometimes makes mistakes.
    \item \textbf{Direction} criterion determines whether the subgoals collectively guide the student towards the correct solution.
    \item \textbf{Misleading information} criterion identifies subgoals containing incorrect guidance, such as non-existent functions or wrong constants. 
    \item \textbf{Granularity} criterion examines if the subgoals are limited to a single action.
    \item \textbf{Idiomatic} criterion evaluates whether the subgoals adhere to the idiomatic practices of the target programming language, in our case --- Kotlin.
\end{itemize}

\subsubsection{Methodology} \label{sec:first_methodology} We conducted two rounds of validation by four experts with up to 5 years of programming and teaching experience.
For each task in the studied Kotlin course, we took the initial state of the solution at the start of solving and ran the first step of our pipeline with the subgoal generation. Then, the experts independently labeled the LLM output for each task, with each being labeled by at least two experts. The labeling of each list of subgoals consisted of deciding whether it satisfies each validation criterion. After labeling, all experts gathered and finalized their decisions during an online meeting, reaching a consensus for each task. Based on the results of the first round, consisting of a \textit{success rate} of each criterion (\textit{i.e.}, the ratio of tasks, for which the generated subgoals satisfied it), we implemented certain changes (described below) and conducted a second, final, round of validation.

\subsubsection{First round of validation} The first round of validating the generation of subgoals was conducted for all coding tasks from all six projects of the course (50 tasks in total). 
The results were generally positive. They demonstrated that the LLM is capable of handling the task of subgoal generation, achieving a success rate of over 70\% for most criteria. However, issues were identified for the following criteria: \textbf{Specifics} (38\% success rate), \textbf{Coding-specific} (52\% success rate), and \textbf{Misleading information} (58\% success rate). Let us consider some specific problems in more detail, and describe what changes we implemented to overcome them. An example of the generated subgoals with these problems can be found in the supplementary materials~\cite{suplementary}.
\begin{itemize}
    \item \textit{Limited recognition of built-in Kotlin functions}. Sometimes the LLM provided an algorithm that does not suggest to use built-in functions. One of the popular ways to fix this problem is to use the retrieval-augmented generation (RAG)~\cite{lewis2020retrieval} technique, which we leave for future work. The same approach can be used to improve the performance on the \textbf{Misleading information} criterion.
    \item \textit{Insufficient or excessive granularity}. During our first iteration, we discovered that the LLM could not accurately determine the size of a subgoal, \textit{e.g.}, merging several logical subgoals into a single big one or providing many subgoals not related to code. We applied several prompt adjustments to resolve this issue:
        \begin{itemize}
            \item Provided the LLM with a description of what a subgoal means~\cite{margulieux2020effect};
            \item Added an instruction to label all subgoals as \textit{code} and \textit{no-code} to then filter all \textit{no-code} ones programmatically.
        \end{itemize}
    \item \textit{Suggesting redundant actions}. The last problem was related to adding extra subgoals that were not asked for in the task, \textit{e.g.}, \textit{test or run your solution}. It is related to the previous one and was solved by adding labels to each subgoal.
\end{itemize}

\subsubsection{Second round of validation} After making improvements, we carried out the second round of validation. We randomly chose 16 out of 50 tasks from the course  (32\%), representing all six projects. The results of the second round showed significant improvements in the criteria that were problematic in the first round. The success rate for \textbf{Specifics} increased from 38\% to 75\%, and for \textbf{Coding-specific} --- from 52\% to 81\%.  Additionally, we observed slight improvements in the \textbf{Misleading information}, \textbf{Independence}, and \textbf{Idiomatic} criteria. The overall success rate for the remaining the criteria was over 70\% as well. 

\subsection{Validating the Generation of Hints}\label{sec:design:hints}

The second and third steps of our approach  are \textit{generating code and textual hints} (see Section~\ref{sec:approach:hints}). This section describes how these steps were validated with experts. We carried out this validation for both types of hints together, because they constitute different levels of representation of the same hint.

\subsubsection{Validation criteria} In order to accurately validate  the quality of the generated textual and code hints, we propose to use twelve criteria, eight of which were taken and adapted from the literature~\cite{keuning2018systematic, roest2024next} and four taken from our experience and preliminary experiments. The eight criteria from the work of Roest et al.~\cite{roest2024next} are as follows.

\begin{itemize}
    \item \textbf{Feedback type} criterion characterizes what kind of feedback is generated, such as if the hint takes into account the knowledge about task constraints or explains to the student the reasons why the hint was proposed. This criterion has several sub-criteria, their detailed descriptions can be found in the original work~\cite{roest2024next}.
    \item \textbf{Information} criterion identifies any additional information in the hints, which could potentially help the student to understand the hint better. 
    \item \textbf{Level-of-detail} criterion defines if the generated hint is a high-level description or a specific bottom-out hint. 
    \item \textbf{Personalized} criterion indicates if the hint refers to the student code and can be a logical extension of the current solution. 
    \item \textbf{Appropriate} criterion generally describes if the hint is a suitable next step, given the current state of the student program and the desired outcome.
    \item \textbf{Specific} criterion checks the size of the hint and whether it is limited to a single step. 
    \item \textbf{Misleading information} criterion indicates if the hint contains misleading information, \textit{e.g.}, asks to use incorrect functions or undefined variables. 
    \item \textbf{Length} criterion for the textual hint is the measurement of the number of words and sentences. We included into the criteria the \textbf{Length} of the code hint as well, which is calculated in terms of the number of added, changed, and deleted lines of code.
\end{itemize}

We introduce four new criteria:

\begin{itemize}
    \item \textbf{Intersection} criterion indicates whether the suggestion in the hint is fully or partially implemented in the student's code. The aim of this criterion is to mitigate cases where hints repeat student's code. 
    \item \textbf{Code quality} criterion indicates whether the generated code is compilable, and does not have common code quality violations such as incorrect parentheses or brackets.
    \item  \textbf{Idiomatic} criterion is designed to check that the generated code uses language-specific constructs, \textit{e.g.}, using the \textit{random} function from Kotlin to choose a random element from a list rather than using generic Java-like code with a loop.
    \item \textbf{Subgoals relevance} criterion checks if the generated hint matches with the list of subgoals proposed at the previous step of the algorithm.
\end{itemize}

The full description of all criteria can be found in the supplementary materials~\cite{suplementary}. 

\subsubsection{Methodology} We conducted two rounds of validation by two experts with up to 5 years of programming and teaching experience. 
To check more varied hints, we collected real student submissions for the studied Kotlin course and used them for hint generation. During each round of validation, we ran the entire hint generation pipeline on each student submission. The labeling procedure was the same as in Section~\ref{sec:first_methodology}, with two experts labeling each generated hint based on the proposed criteria and then reaching an agreement. Similarly, based on the success rate of various criteria, certain changes were implemented, and the second round of validation was conducted.

\subsubsection{First round of validation} The first round of validating hints was conducted on 24 submissions from 18 students across all six projects from the course. We collected the submissions that did not pass the tests, since this was one of the most appropriate places to show the hints. In this round, we only considered 24 submissions, because after labeling them we found that the initial quality of the approach was too low, and we did not need to label any more data. 

\begin{figure*}[t]
    \centering
    \includegraphics[width=\linewidth]{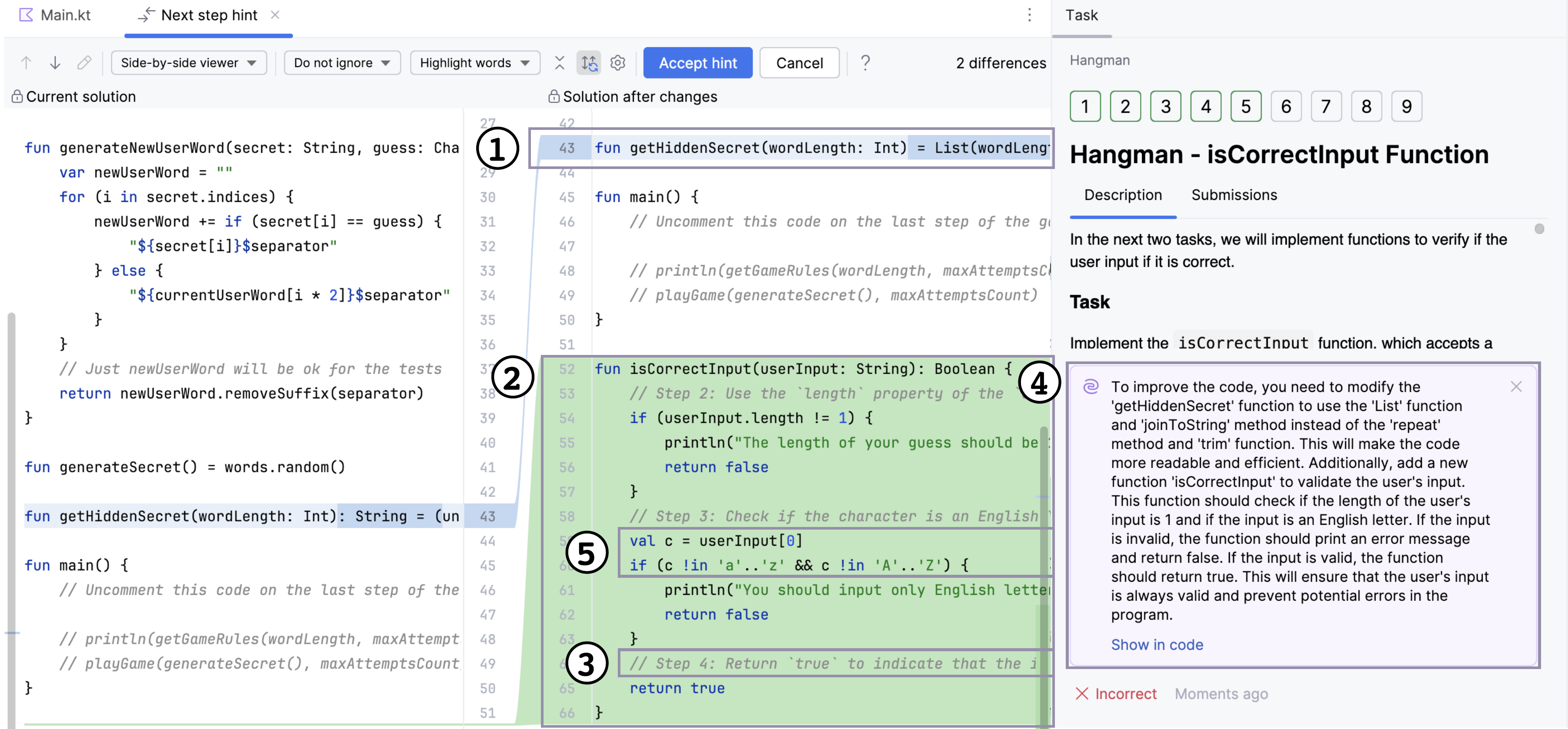}
    \caption{Example problems in the hint: (1) Unintended code modification, (2) Premature disclosure of the full solution, (3) Unwanted procedural comments in code, (4) Excessive detail in texual hint, (5) Limited recognition of built-in Kotlin functions.}
    \label{fig:example_problems_in_hint}
\end{figure*}

While most criteria achieved a success rate of at least 60\%, some had notable issues. The \textbf{Misleading information} had the lowest rate at 27\%, followed by \textbf{Information} at 32\%, and \textbf{Appropriate} at 41\%. 
We highlighted several problems (see examples in Figure~\ref{fig:example_problems_in_hint}): 
\begin{itemize}
    \item \textit{Unintended code modification}. In the generated code, the LLM often modified code that should not have been modified (see \textbf{(1)}). The diff window shows that the proposed hint tries to modify the \textit{getHiddenSecret} function implemented in the previous task. These changes should not be pointed out to the student since that part is already correct. As mentioned in Section~\ref{sec:approach:hints}, we  applied static analysis to mitigate these problems:
        \begin{itemize}
            \item Using static analysis, we defined the list of functions that need to be implemented in the current task and then analyzed the generated code to include only these functions in the hint. 
        \end{itemize}
    \item \textit{Premature disclosure of the full solution}. We discovered that the LLM often ignores the size of the step provided by a subgoal and suggests a large step (see \textbf{(2)}). To fix this problem, we also applied static analysis:
        \begin{itemize}
            \item We developed three general heuristics, each applied to six control structures (see Section~\ref{sec:approach:hints}). This helped us manually control the code hint size and avoid recommending large constructions in one hint.
        \end{itemize}
    \item \textit{Unwanted procedural comments in code}. Sometimes, the LLM added comments into the hint that referred to the previous subgoals from the list of subgoals (see \textbf{(3)}). Since we do not show the entire list of subgoals to the students, this can confuse them. We applied static analysis to fix this problem as well:
        \begin{itemize}
            \item After the code hint is generated, we analyze the generated code and remove all comments. Since the code hint is accompanied by a textual hint, and we manually control the size of the hint, we assume that we do not need to include additional comments in the code.
        \end{itemize}
    \item \textit{Excessive detail in textual hint}. This problem was connected to the size of the textual hint (see \textbf{(4)}). To control the size of the hint, we changed our approach:
        \begin{itemize}
            \item We changed the order in which the hint is produced, generating the code first, and the text second, as was done in the recent work~\cite{kazemitabaar2024codeaid}. 
            \item We changed the prompt for generating the textual hint and removed the word \textit{explain}.
        \end{itemize}
    \item \textit{Limited recognition of built-in Kotlin functions}. Similar to generating subgoals, we noticed that the LLM often ignores built-in Kotlin functions (see \textbf{(5)}). To mitigate this issue, we applied the following change:
        \begin{itemize}
            \item We added a heuristic for short functions (see Section~\ref{sec:approach:hints}), recommending the solution pre-written by the course creator instead of the LLM-generated one for functions not larger than three lines of code. We assume that the course creator uses optimal language constructions in their solution. We chose the threshold of three lines empirically, based on the experience of the experts, and left the detailed experiments for future work.
        \end{itemize}
\end{itemize}

\subsubsection{Second round of validation} After making the improvements, we carried out the second, final, round of validation. Since our goal is to provide hints not only for errors, we extended the data with different stages of students' solutions. For that, we adopted the TaskTracker tool~\cite{lyulina2021tasktracker} for our experimental setup. The tool collects all intermediate changes during the course of solving the given task, allowing us to see different stages of solutions, not only after failing the tests. In total, we collected the data from eight first-year Bachelor's students solving all 50 tasks in the ``Kotlin Onboarding: Introduction'' course. Among the intermediate stages of their solutions, we randomly extracted 48 compilable versions, with some at the start of the solving process and some very close to the end.

The results of the second round of validation showed a significant improvement in the hint quality among various criteria. First, the \textbf{Appropriate} criterion increased from 41\% to 54\%. 
 The final success rate of the \textbf{Subgoals relevance} criterion also stands significantly high at 79\%. 
 Additionally, we observed a slight improvement in \textbf{Misleading information}. Overall, all problems described in the previous sub-section and shown in Figure~\ref{fig:example_problems_in_hint} were largely solved and did not manifest themselves in the final version of the tool.

\section{Evaluation with Students}\label{sec:evaluation}

After the proposed approach was developed and validated with experts, we conducted an evaluation with students using the hint system. The primary objective of this evaluation was to obtain their feedback regarding the usefulness and the ease of use of the provided hints in a classroom. For this evaluation, we used the final next-step hint system described in Section~\ref{sec:approach:internal}, equipped with \gptFour. 

\subsection{Methodology}

The evaluation consisted of two components. Firstly, we conducted the analysis of log data obtained by the adapted TaskTracker tool~\cite{lyulina2021tasktracker}. The tool logs all IDE actions performed when solving the tasks, as well as all internal LLM-related data (prompts, inputs, outputs of all intermediate stages). For this, 14 students from two universities installed the TaskTracker tool and the improved version of the JetBrains Academy plugin with the hint system. Then, they completed at least five out of six projects in the ``Kotlin Onboarding: Introduction'' course~\cite{kotlin-onboarding-introduction}. There were no restrictions placed on using the ``Get hint'' button. After the students completed the projects, we conducted the second part of the evaluation --- a qualitative assessment through a survey that included open-ended questions about their interactions with the hints.

All students agreed to submit their data via the TaskTracker tool. We only collected data from the files that were created for solving the tasks, and anonymized all the personal data.

\subsection{Results}

\subsubsection{How often do students use the hint generation system?} To answer this question, we conducted a quantitative analysis of the TaskTracker data that we collected.

Among all students and all tasks, 191 hints were requested, of which 101 times students also asked to show the code hint (52.9\%). The distribution of requests among projects was not uniform, for example, only 13 hints were requested in the first two basic projects about reading variables and printing them to the console. This behavior was expected, since all of the students were already familiar with the basic programming concepts, which was enough to solve the first two projects without help.
Students only requested the code hint in about half the cases, and in the other half, it was sufficient for them to see only a textual description of a possible next step. These observations match well with the work of Xiao et al.~\cite{xiao2024exploring}, where it is shown that there are differences between students, with some requiring help beyond a short explanation on what to do next with a textual hint.

Out of 101 code hints shown, 65 were accepted by clicking the ``Accept hint'' button (64.4\%). It should be noted that we did not analyze the code snippets before and after the code hint was shown to find cases where students did not accept the hint but applied the same changes manually. This will be done as part of future work, because it might highlight different student behaviour patterns.

Out of 191 next-step hints (both textual and code), 47 were regenerated, \textit{i.e.}, the student requested the hint again without accepting the previous one or changing the code. This highlights the value of using LLMs as opposed to the solution provided by the task creator, since LLMs can generate different suggestions from the same state.

\subsubsection{How do students perceive the proposed hint system?} To answer this question, we conducted a  survey.

\textbf{Reasons to ask for hints.} We asked students about the reasons they had for requesting a hint. The most popular responses were \textit{I was curious} (9 responses) and \textit{My solution did not pass the tests} (7 responses). The options \textit{Didn’t know what to do next} and \textit{Compilation errors} were also indicated by five students. This confirms that the proposed hint system has been used for its purpose, although there is room for improvement in generating hints for compilation errors. One of the students shared an unusual scenario where they completed the entire project using only the suggested hints. This indicates the importance of future considerations on whether the number or the frequency of the allowed hints should be limited.

\textbf{Unclear hints.} To better understand the students' behavior, we asked them about their actions when the generated hint was unclear. Surprisingly, most of the responses (11 out of 14) indicated that the students tried to solve the task themselves, without any extra help such as Google or resources such as ChatGPT. The second popular action (7 responses) was regenerating hints. This behavior of students shows the importance of high accuracy in hint generation, otherwise many students will not trust such a system and will try to solve problems without additional help.

\textbf{Hints representation.} Half of the students (7 out of 14) prefer to see the combination of textual and code hints, while 6 out of 14 noted they would prefer to have code hints even if the textual hints are omitted. This indicates the importance of having several levels of hints, but again raises the questions about limiting the access to them. We will consider these limitations in future work.

\textbf{General feedback.} In general, the proposed hint system was well received by students --- 8 out of 14 indicated they would continue using the proposed system, with most of them being novices with little programming experience. The proposed system is useful for them, because they do not know exactly what to ask ChatGPT or Google. At the same time, more experienced students reported that they lacked conversations in this system as they could not ask specific questions. This indicates the importance of combining approaches such as ours and approaches like CS50.ai~\cite{liu2024teaching} with a chat-based assistant to support students of all experience levels. 
\section{Limitations \& Discussion}\label{sec:threats}

\textbf{User experience study.} In order to design the user interface of the proposed system, we conducted a short interview study with nine students. It is possible that the sample of students is not representative enough or that not all possible UI options were considered. It is also possible that the option most preferred by the students is not the most useful for their learning progress. In the future, it is necessary to conduct more thorough analysis of different potential ways to demonstrate the hints to the students.

\textbf{Evaluation with students.} We conducted the evaluation with only 14 students in a classroom, collecting their usage and feedback about the system. However, we did not conduct a detailed comparison of how students perform with and without the proposed next-step hints and whether the students actually learned from the hints. This constitutes the most important part of our future work.

\textbf{Generalizability.} We implemented the hint system as part of the JetBrains Academy plugin for Kotlin courses. While this limits the range of possible applications, our approach is generally language-independent and can be extended to other courses for other programming languages, as well as other environments such as Visual Studio Code. To support another language, the proposed static analysis techniques should be re-implemented for it, \textit{e.g.}, the size heuristics and the code quality inspections. However, a lot of existing tools and IDEs can provide such functionality.

\section{Conclusion}\label{sec:conclusion}

In this work, we presented a next-step hint system that provides textual and code hints to students. The system combines state-of-the-art LLMs and static analysis to create high-quality next-step hints. The system implements a chain-of-thought approach and consists of three distinct stages: (1) generating subgoals, \textit{i.e.}, a list of steps that need to be taken to solve the current task, (2) generating the code hint that would implement the next subgoal, and (3) generating the textual hint to explain the necessary change in a concise manner. During the generation of code hints, we used static analysis to control their size and code quality. 

We implemented the proposed hint system as part of the open-source JetBrains Academy plugin. We extended the existing interface with a new ``Get Hint'' button that aims to display the generated hint in two stages. First, we show the textual hint and point to the location in the student's code where the changes should be applied. Then, as an optional step, we provide the student with a code hint that can be accepted or rejected. We carried out several rounds of internal validation for each step of the proposed pipeline, which allowed us to improve the quality of the final solution.

Finally, we carried out an initial evaluation with 14 students in a classroom. The students actively used the proposed system to get help, and the results indicated the importance of having different forms of hints for different students. Overall, the proposed next-step hint generation system demonstrates potential, and we plan to incorporate it more fully into the learning process and conduct detailed studies on the value it brings to the students.

\balance
\bibliographystyle{ACM-Reference-Format}
\bibliography{cite}

\end{document}